\documentstyle[11pt,aaspp4]{article}
 
\newcommand{\kms}{$\,{\rm km\,s^{\scriptscriptstyle -1}}$}
\newcommand{\gtsim}{\ {\raise-0.5ex\hbox{$\buildrel>\over\sim$}}\ }
\newcommand{\ltsim}{\ {\raise-0.5ex\hbox{$\buildrel<\over\sim$}}\ }
\def\Msun{\hbox{$\thinspace M_{\odot}$}}
\def\Lsun{\hbox{$\thinspace L_{\odot}$}}

\journalid{}{}
\articleid{11}{14}  

\slugcomment{To be published the Astronomical Journal}
 
\begin{document}

\title{The Formation and Evolution of Candidate Young Globular Clusters 
in NGC 3256}
 
\author{Stephen E. Zepf}
\affil{Department of Astronomy, Yale University, New Haven, CT 06520; \\
zepf@astro.yale.edu}

\author{Keith M. Ashman}
\affil{Department of Physics and Astronomy, University of Kansas,
	Lawrence, KS 66045; ashman@kuspy.phsx.ukans.edu}

\author{Jayanne English}
\affil{Space Telescope Science Institute, Baltimore, MD 21218;
jenglish@stsci.edu}

\author{Kenneth C. Freeman}
\affil{MSSSO, Private Bag, Weston Creek PO, 2611 Canberra, ACT,
Australia; \\ kcf@merlin.anu.edu.au}

\author{Ray M. Sharples}
\affil{Department of Physics, University of Durham, South Road, Durham,
DH1 3LE, UK; \\ r.m.sharples@durham.ac.uk}

\begin{abstract}

        We present images of the recent galaxy merger NGC~3256 obtained 
with the WFPC2 of the Hubble Space Telescope in $B$ and $I$ filters. 
We show that there is a large population of more than 
1,000 compact, bright blue objects in this galaxy within the 
7~kpc $\times$ 7~kpc region studied.
These objects have sizes, colors and luminosities like those 
expected for young Galactic globular clusters with ages ranging from 
a few to several hundred Myr. 
On this basis, we identify at least some fraction of the compact, 
bright blue objects in NGC 3256 as young globular clusters. 
The young cluster system comprises a significant fraction of the 
total luminosity of the galaxy within the region studied -  
$15-20\%$ in $B$ and half that in $I$, indicating a high efficiency 
of cluster formation on a galaxy-wide scale.
In order to determine the properties of this young cluster system,
the selection effects in size, color, and luminosity are carefully
modeled. We find that the intrinsic color distribution is
broad, and there is no significant trend of color with magnitude. 
The combination of the broad range of observed colors and
the lack of a trend of redder colors at fainter magnitudes
can not be fit solely by a broad age distribution
and/or differential reddening, although the latter is
clearly present. The observations  
can be accounted for by either the preferential depletion/destruction
of lower mass clusters as they age, or a very young age 
($\ltsim$ 20 Myr) for the cluster population, comparable to or less
than the dynamical time of the region in which the clusters
are observed. We also find that the luminosity function of the 
young cluster system can be roughly fit by a power-law with an 
exponent of -1.8, with tentative evidence that it flattens at 
faint magnitudes. The clusters are compact in size, with typical
estimated half-light radii of five to ten parsecs, but there 
is no obvious cut-off for larger radii, and only a shallow trend 
of size with luminosity. We discuss the implications
of these results for models of the formation and dynamical
evolution of globular clusters, as well as for interpretation
of the properties of older globular cluster systems.

\end{abstract}


\section{Introduction}

	Globular clusters have long been recognized as 
excellent fossil records of the formation history of their
host galaxies (Ashman \& Zepf 1998 and references therein).
They also provide critical testbeds for the study of stellar
evolution and stellar dynamics. 
However, the formation process of globular clusters themselves
is not well understood. 
One hypothesis is that merger-induced starbursts are favorable 
environments for globular cluster formation (Schweizer 1987, 
Ashman \& Zepf 1992). Ashman \& Zepf (1992) specifically predicted 
that Hubble Space Telescope (HST) images of gas-rich mergers 
would reveal young globular clusters, readily identifiable through
their very compact sizes, high luminosities, and blue colors.
This prediction has been dramatically confirmed. 
Initial discoveries of compact, bright, blue star clusters
in HST images of the peculiar galaxy NGC~1275 by 
Holtzman et al.\ (1992) and in the well-known galaxy merger 
NGC~7252 by Whitmore et al.\ (1993), 
have been followed up by the observations of similar
objects with the characteristics of young globular clusters 
in a large number of starbursting and merging
galaxies (see list in Ashman \& Zepf 1998).
The identification of these compact, bright, blue objects
as young star clusters has been confirmed by ground-based 
spectroscopy in several systems (e.g.\ Schweizer \& Seitzer 1998, 
Brodie et al.\ 1998, Zepf et al.\ 1995, Schweizer \& Seitzer 1993).
There are even possible mass estimates from high resolution
spectroscopy of a few nearby examples (e.g.\ Ho \& Filippenko 1997).

	These observations provide significant support for the 
idea that globular clusters form in galaxy mergers and strong 
starbursts. They also suggest that globular cluster formation 
may be a regular part of the starbursting process.
The empirical evidence for globular cluster formation
in these environments is broadly consistent with the hypothesis
that globular clusters primarily form in mergers and starbursts
rather than in other sites.  Other globular cluster formation
scenarios appear to have difficulties accounting for the
observational properties of globular cluster systems 
(e.g.\ Ashman \& Zepf 1998, Harris 1996 and references therein). 
In particular, correlations between cluster and host galaxy properties 
and the absence of dark matter are problematic for primordial
globular cluster formation models (e.g.\ Peebles \& Dicke 1968, 
Rosenblatt et al.\ 1988). Similarly, thermal instability models 
for globular cluster formation (e.g.\ Fall \& Rees 1985)
appear to be unable to account for the absence of a correlation 
between globular cluster metallicity and mass along with the 
high metallicities of typical globular clusters ([Fe/H] $> -1$).

	The discovery of young globular cluster systems in
nearby starbursts and galaxy mergers opens up the possibility 
of more detailed, empirical studies of the formation and evolution
of globular clusters. One of the questions that remains to be 
answered is the efficiency with which globular clusters form 
in starbursts and mergers. This efficiency is critical for
determining if most or all globular clusters can form in
merger-like conditions. The efficiency can also constrain 
models of the formation the clusters themselves. For example, models 
in which globular clusters form as cores in much larger clouds 
predict low efficiencies, while models in which typical molecular 
clouds are compressed may be more efficient.

	A closely related question is the dynamical evolution of
globular cluster systems. Most studies to date have concentrated
on developing theoretical models and matching these to the
properties of old globular cluster systems that have undergone
evolution over most of a Hubble time. Attempts to infer the
initial population and the effects of evolution from the remnant
population are difficult. Observations of young cluster systems 
can provide valuable input into the initial conditions and early 
dynamical evolution of globular cluster systems. This is not only 
true of the mass (luminosity) function, but also of the 
radii and densities with which the clusters form.

	The efficiency of globular cluster formation in mergers
and of the dynamical evolution of globular cluster populations 
also have significant implications for the use of globular 
cluster systems as fossil records of the formation history 
of their host galaxies. For example, Ashman \& Zepf (1992) 
predicted that if elliptical galaxies form by mergers, they 
should have two or more populations of globular clusters. 
One of these populations originates from the halos of the 
progenitor spirals and is therefore spatially extended and 
metal-poor, while the other forms during the merger and is 
thus more spatially concentrated and metal-rich.
This prediction of multiple populations in the globular cluster
systems of ellipticals formed by mergers has now been confirmed 
in many cases (e.g.\ Ashman \& Zepf 1998 and references therein).
However, it has not yet been clearly demonstrated that 
the efficiency of cluster formation in galaxy mergers is
sufficient to account for the metal-rich globular cluster 
population observed in elliptical galaxies. Furthermore, although
there are strong theoretical arguments that the mass function
of globular cluster systems evolves significantly over
time to resemble the log-normal mass function of old
globular cluster systems (e.g.\ Gnedin \& Ostriker 1997,
Murali \& Weinberg 1997a), this evolution has not
been demonstrated observationally.

	The goal of this paper is to address the questions
of the formation and evolution of globular clusters through
the study of the galaxy merger NGC~3256. The HST observations
on which this study is based and the analysis of these data 
are presented in $\S 2$. The resulting sample of a large number 
of compact, bright, blue objects in NGC~3256 is examined in
detail in $\S 3$. This section includes the 
determination of the relationship between the magnitudes, colors, 
and radii of the young cluster sample and the luminosity function.
The implications of the results for the formation efficiency and 
dynamical evolution of globular cluster systems are discussed
in $\S 4$, and the conclusions are given in $\S 5$.

\section{Observations and Data Reduction}

\subsection{Target Galaxy}

	We utilized HST and WFCP2 to obtain high resolution images of
the galaxy NGC~3256. This galaxy was selected for our program because 
it has long been identified as a galaxy merger (e.g.\ Toomre 1977)
and is fairly nearby, with $cz_{\odot} = 2820$ \kms\ (English et al.\ 1999) 
which places NGC~3256 at a distance of 37 Mpc for 
H$_0$ = 75 km s$^{-1}$ Mpc$^{-1}$. As shown in Figures 1 and 2 
(plates 1 and 2), the central region of NGC~3256
has star forming knots, dust lanes, and loops,
along with a more extended, smoother component. In the radio
continuum and at $2.2 \mu$ there appear to be two nuclei
separated by about $5''$, or about 1 kpc (Norris \& Forbes 1995,
Kotilainen et al.\ 1996, Doyon, Joseph, \& Wright 1994). 
Tidal tails can also be seen in the optical
images in Figure 1, and have been shown to extend out to $\sim$ 
50 kpc in HI (English et al.\ 1999).
Toomre (1977) placed it in the middle of his sequence of 
disk galaxy mergers, 
suggesting that it is dynamically older than the NGC~4038/4039 system 
(the Antennae), but younger than NGC~7252. 
Of the eleven mergers on the Toomre list, NGC~3256 also
has the most molecular gas ($1.5 \times 10^{10}$ \Msun, Casoli et al.\ 1991,
Aalto et al.\ 1991, Mirabel et al.\ 1990), and is the
brightest in the far-infrared ($L_{FIR} = 3 \times 10^{11}$ \Lsun,
Sargent et al.\ 1989).

\subsection{HST Observations}
	
	The WFPC2 images of NGC~3256 were obtained with the
Planetary Camera (PC) centered on the galaxy. At a distance of 37 Mpc, 
each PC pixel is 8 pc, and the PC covers a total of 7 kpc $\times$ 7 kpc, 
encompassing the starburst region identified in previous studies. 
The PC data centered on NGC~3256 are the subject of this paper. 
The data at larger radii will be discussed in future papers.

	We imaged NGC~3256 in the F450W and F814W filters.
Two equal exposures were obtained through each filter, with 
total exposures times of 1800s in F450W and 1600s in F814W.
For each filter, the two exposures were combined utilizing
a cosmic-ray rejection routine kindly provided by Rick White.
As a check on this procedure, we also performed the more standard
CCREJECT task in STSDAS on the images in each filter, and then 
set a strict criterion for matching the object lists
between the two filters. The final results were very similar 
to those produced by White's routine (c.f.\ Schweizer et al.\ 1996,
Miller et al.\ 1997). A visual examination
of the few differences favored the results of White's routine,
so we used these combined images for further analysis.
In any case, the number of compact sources observed is far 
greater than any possible residual defects. 
The resulting combined images are shown in Figure 2 (Plate 2).


\subsection{Cluster Identification}

	A wealth of blue, compact objects is revealed in the
HST images shown in Figure 2 (Plate 2).
In order to determine the magnitudes and sizes of 
the compact objects discovered in the HST images,
we first used the DAOFIND task in IRAF to identify objects.
This task convolves the image with a Gaussian kernel,
finds the best fitting Gaussian function at each point, and
then searches for density enhancements which are both
greater than a given threshold value and the 
brightest density enhancement in a localized region determined
by the width of the Gaussian kernel.
For this analysis, we set the FWHM of the Gaussian to be
2.8 pixels, which is the apparent width expected for 
an object with a true FWHM of roughly 2 pixels.
We also applied broad cuts with the DAOFIND sharpness
and roundness criteria to eliminate a few extremely diffuse or
sharp features. 

	There are two notable effects of identifying objects in
this standard way. One is that it introduces a selection bias
against objects significantly larger than the smoothing kernel. 
This is a direct result of the search for density enhancements 
on a given scale. Although not an issue if all of the objects
are unresolved or marginally resolved, this selection effect
needs to be accounted for in studies of the distribution of
object sizes. A second aspect of DAOFIND is that the threshold is 
defined globally. Several other globular cluster searches 
have been performed using a local threshold, rather than a 
global one (e.g.\ Kundu et al. 1998, Carlson et al.\ 1998,
Miller et al.\ 1997). Although this has the advantage of giving 
a uniform number of false detections over the image, it does 
so at the cost of producing a non-uniform magnitude limit 
across the frame. As this is critical for our purposes, 
we retain the global threshold, and simply set it high
enough that the probability of a spurious source in
regions of high background (noise) is negligible.
Perhaps the most critical aspect is that the detection
algorithm is well-understood, and can be run on a variety
of artificial datasets to explore the success with which is recovers
objects of various luminosities, colors, and sizes.

\subsection{Cluster Photometry}

	The next step is to determine the magnitudes of the
identified objects. Because of crowding, variable background, and
signal to noise, it is not possible to determine the brightness
profile of the objects out to large radius. Therefore we
perform aperture photometry from one to several pixels in
radius, and correct these modest apertures to total magnitudes. 
If the objects were unresolved, the aperture
correction to total magnitude would be straightforward,
as the HST point spread function (psf) is reasonably well-understood. 
Moreover, an aperture of several pixels incorporates the majority 
of the light from an unresolved source, even in the PC, 
so the overall correction is not a large one. However, 
the objects we detect in NGC~3256 are resolved, as expected 
for objects with sizes like those of Galactic globular clusters 
at the distance of NGC~3256. In this case, the aperture
corrections depend on both the psf and the intrinsic
radial profile of the object.

	There is a limited amount of spatial information in the
surface brightness profile within the few pixels radius
out to which it can be reliably measured. Therefore, if a form
of the profile is assumed, the radial scale of that profile can
be determined by the difference between magnitudes at small
radii. A Gaussian shape for the intrinsic profile of the
clusters has been adopted in most previous work (e.g.\ Whitmore et al.\ 1993, 
Whitmore \& Schweizer 1995, Schweizer et al.\ 1996). 
In order to facilitate comparison with these studies, we
also adopt a Gaussian profile for the clusters, 
although we note that this will tend to underestimate 
the total magnitude if the clusters follow a more
extended profile, such as a modified Hubble law
(e.g.\ Holtzman et al.\ 1996, Carlson et al.\ 1998, Ostlin et al.\ 1998).

	In detail, we determine the size of each object
by comparing the difference between magnitudes within 
apertures of one and three pixels ($m_3 - m_1$) 
to a table of values for a wide range of
Gaussian profiles convolved with the HST psf given by the 
TINYTIM software (Krist 1993). 
This is done in both the B and the I filters,
and the resulting intrinsic FWHM is taken as the average
of the two. We tested this procedure by using I-band (F814W) 
observations of unsaturated stars in the globular cluster
Omega Cen as the basis for the HST psf. This psf gives the
same inferred sizes as the TINYTIM psf when the intrinsic 
input FWHM is greater than about 0.5 of a PC pixel.
For objects with intrinsic sizes less than 0.5 of a PC pixel, 
a slightly ($\sim 10\%$) smaller size is inferred with the 
Omega Cen psf than with the TINYTIM psf. This difference 
in inferred size due to different psfs is much smaller than 
the uncertainty introduced
by the assumption of a form for the intrinsic surface brightness 
profile of globular clusters with only a single free parameter.
Therefore, we adopt the TINYTIM results for the remaining analysis. 
We note that although the absolute errors in total magnitudes
and half-light radii due to the requirement of an assumed
profile form for the clusters in the sample can be significant, 
they should give good {\it relative}
sizes and total magnitudes, providing the clusters are roughly
similar to each other in structural characteristics.
Moreover, any systematic error is likely to affect both filters,
so that the colors will be mostly unaffected.

	The total magnitude of each object is determined
by applying the aperture correction from the magnitude within
an aperture of 2 pixels in radius to a total magnitude,
appropriate for the measured
size for each object. This aperture correction for an object
of typical size in our sample is roughly 0.85 magnitudes in B
and 1.0 magnitudes in I. It compares to 0.25 magnitudes in B
and 0.48 in I for completely unresolved objects. These differences
emphasize both that our objects are significantly resolved, and
that the colors are largely unaffected by the correction
to total magnitudes (cf.\ Holtzman et al.\ 1996). 
We correct the magnitudes and colors of the objects for Galactic 
reddening using the dust maps of Schlegel, Finkbeiner, \& Davis (1998).
These give $E(B-V)= 0.12$, and extinction 
in our HST filters of $A_{F450W} = 0.47$ and $A_{814W}=0.23$.
The absolute photometric calibration to
the standard B and I$_C$ system is achieved using the 
Holtzman et al.\ (1995) zero points for the F450W and F814W filters.

\section {Analysis}

	The color, magnitude, and sizes of the objects detected
in NGC~3256 are plotted in Figure 3 \footnote{The positions, magnitudes, 
colors, and sizes of the objects found in NGC 3256 are also given 
in Table 1, available either in the electronic journal or from the 
first author.}. 
Several features of the cluster system of this galaxy merger are apparent 
in this diagram. One is the very large number of bright star clusters 
in this galaxy, approximately 1,000 with $M_{B} < -9$ inside of
the central 7 kpc $\times$ 7 kpc. These objects account for 
approximately $19\%$ of the B light and $7\%$ of the I light 
within this region. The star clusters generally have blue
colors. The bright magnitudes and blue colors are indicative
of massive star clusters at young ages.
For reference, Figure 4 shows the prediction of two
stellar population models for the color and magnitude 
of a $2 \times 10^5$ \Msun\ globular cluster as a function of age.
The clusters are also compact like globular clusters, with typical
sizes of $\ltsim 10$ pc. 
Only a few of the 1,000 objects are expected to be compact background
galaxies or foreground stars, based on similar analyses of
blank fields, such as the Hubble Deep Field.

	In order to address the true distribution of the population
in color, size, and magnitude, simulations of artificial datasets
are required to calibrate the detection procedure. For example,
the absence of objects with very faint B magnitudes and blue B-I
colors may be caused by the detection limit in the I band.
Similarly, the absence of large, faint clusters may also
be possibly due to selection effects. Therefore,
we created a grid of artificial objects with a range of magnitudes
in each bandpass, and a range of sizes for each magnitude. 
By creating a full grid of artificial stars in both bandpasses,
we can address the issue of any effect of incompleteness in
B or I on the color distribution at faint magnitudes.
The difference between input and output magnitudes also
provides a calibration of the effect of ``bin jumping''
when constructing luminosity functions.
Similarly, by incorporating a range of sizes in the artificial
star tests, we can address the question of the intrinsic
size distribution of the cluster population. This study is 
the first in which all of these effects have been modeled.

	We can then test the consistency of various models
of the luminosity, color, and sizes of the candidate globular
clusters in NGC~3256 against the observations. Specifically,
we create model data sets with various combinations of
luminosity functions, color-magnitude relations, and
luminosity-size relations. For the luminosity function
and luminosity size-relation, we adopt a power-law
form, while we use a linear relation between color and magnitude.
The intrinsic widths of the color and size distributions are 
drawn from the data at bright magnitudes where they are unaffected
by selection. Predictions for observables for each model
are made by convolving the model with the
selection functions derived from the simulations
described above.

	We compare these predictions of various models to
the luminosity function, color-magnitude
relation, and luminosity-size relation observed for the
candidate young globular clusters in NGC~3256 (Figure 3).
A model is considered to fit the color-magnitude and luminosity-size
relation if the linear regression of these parameters is
statistically consistent with the data. To insure that
the results are not dependent on the use of a linear fit, 
we also compare the median colors and sizes as a function
of magnitude of the models to the observations, with the
uncertainties in the medians of the data determined via bootstrapping. 
For the luminosity function, the goodness of the fit is determined 
using the double-root-residual test (Ashman, Bird, \& Zepf 1994). 
We also test the effects of changes in
any one of the underlying distributions on all of the observed
properties, as they are not decoupled from each other.
For example, an underlying
luminosity function can be flat, but if the clusters are
smaller at faint magnitudes, they will be easier to detect,
and the luminosity function will appear to rise at faint
magnitudes. Similarly, a trend of color with magnitude can
also give rise to an apparent luminosity function different
than the underlying one if different color clusters are
detected with different efficiency at faint magnitudes.

	We find that the best fitting model cluster
population has little or no correlation between
luminosity and color (B-I independent of $M_{B}$), 
a shallow correlation between
luminosity and radius ($r \propto L^{0.07}$),
a power-law luminosity function $N(L) \propto L^{-1.8}$.
This best fitting model is shown in Figure 5.
The statistical uncertainties on the parameters of
the underlying cluster population are roughly 
0.05 in the slope of the magnitude-color relation, and
0.1 in the exponent of both the radius-luminosity
relation and the luminosity function. We discuss 
the magnitude-color relation, luminosity-size relation, and 
luminosity function individually in more detail below.

\subsection{Color-Magnitude Relation}

	The broad color distribution and absence of a strong 
relationship between color and luminosity places strong constraints 
on the nature and evolution of the young cluster system in NGC~3256.
In order to produce the observed range of colors, either
differential reddening by dust or a range of ages is required.
However, both reddening and age generally produce fainter magnitudes 
for redder objects (see Figure 4).
This is not observed in NGC~3256,
as shown by the similar color-magnitude diagrams of the data 
(Figure 3) and a simulated data set with no relationship between
color and magnitude (Figure 5).
A quantitative measure of the close agreement between
the observations and a simulated data set with no relationship
between color and magnitude is the similar
median (B-I) colors as a function of B magnitude, which are
given in Table 2. In contrast, a simulated data set with an 
intrinsic slope of (B-I) $\propto 0.5$B, like that expected for
a standard reddening law, gives a much steeper relation between 
(B-I) color and B magnitude, as shown in Table 2.
Similar results are obtained using other robust measures of
the average color as a function of magnitude.
The fundamental result is that we are unable to account for
the broad range of observed colors solely by differential reddening
or a broad age distribution because there is no intrinsic 
trend of redder colors with fainter magnitudes.

	The observations can be accounted for in two ways.
One possibility is that the young cluster system in NGC~3256
has an age distribution up to several hundred Myr {\it and} 
low mass clusters are preferentially destroyed over this timescale. 
In this way, the typical luminosity of the older, redder cluster
population will not become much fainter than the younger bluer, 
cluster population because the younger population will have 
more low-mass clusters. A modest amount of reddening may also
be required to produce the colors of the reddest clusters, but
not so much that a strong color-luminosity trend is produced.

	The lack of a strong color-luminosity relation can 
also be accounted for if most of the clusters are very young.  
At ages up to $\sim 10$ Myr, stellar population
models predict a fairly broad range of colors 
with little change in B luminosity (see Figure 4). 
This effect is due to red supergiants, and is
stronger in the Leitherer et al.\ (1998) models
than the Bruzual \& Charlot (1998) models because of
the increased presence of red supergiants in the former models.
As in the previous case, some reddening may be required
to produce the reddest clusters, but reddening
can not be the primary determinant of the cluster colors, 
or a color-magnitude relation would be introduced, contrary
to the observations. The critical aspect of the possibility
that red supergiants at young ages account for much of the 
observed color spread is the requirement of a very young age 
for the system as a whole because all of the models begin 
to produce a significant color-luminosity trend after $\sim 10$ Myrs.

	Both the destruction and red supergiant hypotheses 
can account for a broad color distribution without a strong 
color-luminosity relation. An additional constraint is that 
the range of ages in the young cluster population would not 
be expected to be less than the dynamical time of the region 
in which they formed. Adopting a radius of 3 kpc for the region 
in which the young clusters are found, and a typical velocity 
of $v \approx 150$ \kms, we find a dynamical time of 20 Myr. 
Thus the very young age hypothesis is marginally inconsistent 
with the requirement that the cluster population not form on 
a timescale shorter than the dynamical time.
Other explanations for the absence of a strong 
color-luminosity trend are strongly constrained by the
large observed color spread. For example, a model in which
younger clusters are more heavily reddened than older
clusters can give a weak color-luminosity trend. However, 
it does so at the cost of narrowing the color spread, and 
therefore fails to account for the observations.

\subsection{Luminosity-Radius Relation}

	A second observation of relevance for models of the 
formation and evolution of young star clusters is the 
luminosity-radius relationship. The NGC~3256 young star 
cluster system has a very shallow relationship between 
radius and luminosity, roughly $r \propto L^{0.07}$.
The relationship of radius with mass is likely to be similar 
to that with luminosity. This follows from the absence of a 
correlation between color and luminosity which suggests the 
mass-to-light ratio is mostly independent of luminosity. 
Thus, cluster mass is likely to be fairly independent
of radius. 

	A weak correlation between mass and radius
has significant implications for the formation and
evolution of globular clusters.
Clouds in hydrostatic equilibrium follow the relationship
$ r \propto M^{1/2} P^{-1/4}$ (e.g.\ Ashman \& Zepf 1999,
Elmegreen 1989). The shallow observed correlation
between radius and mass therefore suggests that
higher mass clusters form at higher pressure.
If confirmed, this will play a significant role in
developing models of globular cluster formation.

	A shallow relationship between mass and radius
also suggests that on average low-mass clusters are formed 
with lower density and are less bound than higher mass clusters.
Therefore, they will be more susceptible to destruction
by mass loss at early ages, and through tidal shocks 
over time. This result is an important input into 
determinations of the effect of dynamical evolution on 
the mass function of clusters, as discussed in the previous 
and the following subsections. In particular, studies
of the dynamical evolution of globular cluster systems
must adopt some relation between radius and mass for
the initial cluster population (e.g.\ Ostriker \& Gnedin 1997
and references therein). Without data from young clusters, 
this relation has been based on observations of old cluster 
populations, whose properties may have already been altered by
dynamical evolution. Thus, observations of the radii
of young cluster systems are an important part of the
study of dynamical evolution of globular cluster systems.

	These conclusions regarding the mass-radius relationship
require confirmation, as the clusters are only marginally
resolved in the present data. We are able to recover differences
in the sizes of the objects, given a form for the radial profile
of the cluster. The inferred mass-radius relationship for the cluster 
population should not be sensitive to the form of the profile
adopted because it is only based on relative values of
cluster radii. Therefore, our result is not likely
to be sensitive to the specific cluster profile chosen,
although it will be affected by any systematic changes in the shape
of the profile with cluster mass.

\subsection{Luminosity Function}

	The luminosity function of the NGC~3256 cluster system
has a best-fitting power with slope about $-1.8$, with tentative
evidence that it flattens at faint magnitudes. This slope is similar 
to that found in other young globular cluster systems in galaxy 
mergers (e.g.\ Whitmore \& Schweizer 1995, Schweizer et al.\ 1996, 
Miller et al.\ 1997, Carlson et al.\ 1997) and also to that 
for populous clusters in the LMC (Elmegreen \& Efremov 1997, 
Elson \& Fall 1985).
The most notable difference between the observed luminosity
function and a power-law convolved with observational selection
is that the data appear to be flatter at the faint end than 
the model. In order to assess the statistical significance
of this difference, we utilized a double-root-residual
test (Ashman et al.\ 1994). The test indicates
the difference is significant at about the $2.5\sigma$ level.
However, given the uncertainties modeling the selection
at these faint levels, any deviation from a power-law
slope is tentative. Although the luminosity function of the 
NGC~3256 cluster system is now roughly a power-law, both the 
luminosity-color and luminosity-radius relation suggest that it 
is likely to evolve significantly over time. This is also expected on
theoretical grounds (e.g.\ Gnedin \& Ostriker 1997, Murali \& Weinberg 1997a).
A comparison of the observations to these theoretical expectations
is presented below.

\section {Discussion}

	HST imaging of the galaxy merger NGC~3256 has revealed
a large population of objects with the bright luminosities,
blue colors, and compact sizes expected of objects like Galactic
globular clusters at young ages. In this section, we explore
in more detail the relation of young clusters observed in starbursting
and merging galaxies like NGC~3256 to old globular clusters, 
and to implications these observations have for models of the 
formation of globular clusters.

\subsection{Cluster Mass Function and Dynamical Evolution}

	The power-law luminosity function is the one feature
of the NGC~3256 young cluster system that is decidedly different 
from that of old globular cluster systems around galaxies like
M87 and the Milky Way, which have lognormal luminosity functions.
Although the masses and sizes inferred for many of the young 
clusters in NGC~3256 are similar to those of old globular clusters
(e.g.\ Figures 3 and 5), 
the difference in the shape of the luminosity function has long 
been used to argue that the clusters in mergers and starbursts 
are fundamentally different than in the old systems
(e.g.\ van den Bergh 1995). However, it has also long been realized
that dynamical evolution can significantly alter the mass function
of clusters systems over a Hubble time (e.g.\ Fall \& Rees 1977,
Gnedin \& Ostriker 1997, Murali \& Weinberg 1997a, 1997b,
Ashman \& Zepf 1998 and many references therein).

	We therefore consider whether observations of the mass
and luminosity functions of cluster systems over a range of ages 
can be consistently accounted for by dynamical evolution. 
The two long-term dynamical processes that may 
be relevant for determining the shape of the globular cluster mass 
function are evaporation and tidal shocking. These have the following
scalings between the lifetime of a cluster and its mass and radius 
$ t_{evap} \propto M^{1/2} ~ r^{3/2} $, and $ t_{sh} \propto M ~ r^{-3} $.
The timescale for tidal shocks also depends on the cluster orbit 
and galaxy potential. As long as these are similar in the
galaxies being compared, the scaling is independent of these parameters.
Therefore, we concentrate on the scaling of the destruction
timescale with cluster mass and radius.

	In $\S 3.2$, we found that cluster mass and radius may
be only weakly dependent on one another with $r \propto L^{\sim 0.07}$.
With this result, the equations above give a timescale for 
destruction that scales as roughly $M^{2/3}$ for both 
evaporation and tidal shocking.
These give relative timescales for destruction of clusters
as a function of their mass.
An absolute timescale can be placed on these scalings
through the known turnover mass and age of the globular cluster
systems of galaxies like the Milky Way and M87. This allows us
to determine the cluster mass scale that is undergoing destruction 
at the ages corresponding to young cluster systems observed in 
galaxy mergers. In this way, we find that at an age of 100 Myr, 
the characteristic mass scale set by dynamical processes is 
$10^{-3}$ of that for the globular cluster systems of the 
Milky Way and M87.
At 500 Myr the corresponding number is about $10^{-2}$ of 
the characteristic mass scale of old globular cluster systems.
This calculation is also based on the simplifying assumption
that the dynamical processes are independent of one another,
which is not strictly true. However, the numbers given above
are not likely to be dramatically wrong as long as the
observation that $M$ and $r$ are mostly independent holds.

	If long-term dynamical processes are responsible for evolution 
of the mass function, the above calculation suggests the ``turnover'' 
in the mass function in young globular cluster systems will occur 
at much smaller masses than in older globular cluster systems 
like that in the Galaxy. These small masses make such a turnover 
very difficult to observe directly in young systems. Specifically, 
the above calculation suggests that clusters with masses of 
roughly $1\%$ of the current turnover mass are being destroyed 
at the ages typical of young globular cluster systems, 
such as the NGC~3256 system studied here, as well as the 
NGC~1275 system (Carlson et al.\ 1998) and the NGC~7252 system 
(Miller et al.\ 1997). This corresponds to clusters five magnitudes 
below the current turnover of the globular cluster luminosity function.

	In order to calculate the detectability of a dynamically
induced turnover in young cluster systems, the expected brightening
of young clusters must also be accounted for. Stellar populations models
(e.g.\ Bruzual \& Charlot 1998) predict that old objects of this 
magnitude were about 5.0 magnitudes brighter in B and 3.7 in V 
at the ages inferred for NGC~3256 from their colors. The corresponding 
brightening  for the slightly older NGC~1275 and NGC~7252 cluster systems 
we will also study is about 4.2 magnitudes in B and 3.1 in V. 
The predicted brightening in these young cluster systems relative 
to $\sim 13$ Gyr old populations is dependent primarily on the mass 
function between several \Msun\ and slightly less than one \Msun. 
The adopted numbers are for a Salpeter (1955) slope of $x =1.35$ 
are somewhat less for flatter mass functions. 

	The net result is that in young cluster systems observed
in the B-band, such as NGC~3256, the observations must reach
absolute magnitudes around the current turnover luminosity
in old globular cluster systems, roughly $B = -6.8$. As can
been seen in Figure 3, the observations fail by several
magnitudes to reach such faint levels. In fact, it will be 
difficult to obtain reliable cluster counts at such magnitudes 
because of potential confusion with individual bright stars, 
which can have similar luminosities.
Analyses of the NGC~1275 and NGC~7252 datasets give similar
conclusions that the highest mass scale at which dynamical 
evolution is expected to be effective is well below the 
observational limit. In particular, for these $\sim 500$ Myr old
systems, the observations would have to reach limits
of about $B \gtsim -6$ and $V \gtsim -5.5$, while the
observations are limited to about two magnitudes brighter.

	The luminosity functions of known cluster systems are 
therefore consistent with a model in which globular clusters 
form with a power-law mass function, and evolve through well-known 
dynamical processes to have the log-normal mass function 
observed in old systems such as the Milky Way and M87. 
Testing this hypothesis by directly looking for the turnover 
in young cluster systems is difficult because the mass scale 
is predicted to be extremely small. This work suggests that 
searches for turnovers in globular cluster mass functions 
might be more profitable in more intermediate-age systems 
in which the predicted mass scale may be accessible to observation.

	In order to improve the predictions for the dynamical
evolution of young globular cluster systems, better constraints
on the initial mass-radius relation are required. Studies of young 
cluster systems are critical for providing these constraints because
these systems have not undergone as much dynamical evolution and 
will more accurately reflect the relationship between mass and 
radius at the time the young clusters formed. This relation plays
a major role in the dynamical calculations. Specifically,
the timescales for evaporation and tidal shocking depend
on both mass and radius. If the relationship between mass
and radius is stronger than it appears to be in our
observations of the NGC~3256 system, the timescales for
disruption of clusters of different masses will be affected.
For example, if $r \propto M^{1/2}$, then $t_{evap}$ will
have a very steep dependence on mass, and $t_{sh}$ will
be shorter for larger masses. In this case, the mass
scale for evaporation at 100 Myr would be only about 
a factor of 5 less than that after 10 Gyr, and only
a factor of 3 less at 500 Myr compared to 10 Gyr.
However, the tidal shocking would then be inversely
correlated with mass, so both high and low mass clusters
would be destroyed and the observational predictions
in this case are less clear.

	Perhaps the safest conclusion then from our
analysis of the formation and destruction of globular
clusters is that information on the radii of the clusters
as well as their mass function is critical for determining
the effects of dynamical evolution and comparing of mass 
functions of young and old cluster systems. The cluster radii
and mass-radius relation also have significant implications 
for models of the formation of young globular clusters.
Further exploration of observational constraints on the 
initial mass-radius relation is a promising route for
future studies.

\subsection {Efficiency of Compact, Massive Cluster Formation}

	The census of compact, young star clusters in NGC~3256
allows the determination of the efficiency with which clusters
form. The simplest characterization of efficiency is the
fraction of new stars formed that are in dense star clusters.
As described in $\S 3.1$, this fraction is about $20\%$
based on the percentage of blue light that is in dense
star clusters. This value is similar to that observed 
in smaller starburst regions by Meurer et al.\ (1995).
It is somewhat larger than the fraction of blue light 
in dense star clusters observed in the older NGC~7252 system, 
suggesting that either the  
formation efficiency peaks near the peak of the starburst
or the dynamical evolution removes some of the clusters
on the timescale of the age of the NGC~7252 system.
A critical aspect of these NGC~3256 data is that they
demonstrate that high formation efficiency can occur
over a large area (7 kpc $\times$ 7 kpc) encompassing
much of a $\sim L_{*}$ galaxy. These data also indicate that
the fraction of stars that form in these dense star clusters
is not negligible, and suggest that this is an interesting
mode of star formation. 

	A second way to characterize the efficiency of
cluster formation is to compare it to the total amount of 
gas available. This is less straightforward than comparing
the fraction of light, but connects more directly to current
theoretical models. The total amount of molecular gas
inferred from CO observations and standard assumptions
regarding the conversion to H$_2$ mass is $1.5 \times 10^{10}$ \Msun\
(Casoli et al.\ 1991, Aalto et al.\ 1991, Mirabel et al.\ 1990).
The total mass in the young cluster system can be estimated
from the color of each object, and a stellar populations model 
that allows color to be converted to age and mass-to-light ratio.
The observed luminosity can then be converted into mass. 
Applying this procedure individually to each cluster in
the NGC~3256 cluster sample and summing up
gives a total mass in the young cluster system of 
$6 \times 10^{7}$ \Msun. This is based on taking all 
objects with $(B-I) < 1.5$ and inferred sizes less than 15 pc,
and using the Charlot-Bruzual stellar population models
with a Miller-Scalo initial mass function. A Salpeter 
initial mass function would increase the mass estimate 
by approximately $50\%$. Reddening of observed objects is less
of a factor because its effects on the mass estimate cancel
out to first order. Specifically, reddening makes the objects
fainter, decreasing the apparent luminosity, but also
redder, increasing the inferred mass-to-light ratio,
so in the end the mass estimate is similar. For example,
the mass estimate for the NGC~3256 cluster system is
only about $30\%$ greater for an internal reddening of
$A_B = 1.0$ compared to $A_B=0.0$. An additional effect
is that some clusters may be extincted beyond detection.

	The resulting efficiency for the formation of massive,
compact clusters computed in this way is then about $0.5\%$. 
This is a lower limit in the sense that cluster formation
is ongoing and there is plenty of gas mass left from which to
form more clusters. If clusters continue to make up $\sim 20\%$ 
of the stars formed, the total cluster mass fraction at the
end of the starburst will be closer to this value.
Therefore, the observations place the formation efficiency
between $0.5-20\%$ depending on how exactly efficiency
is defined and what happens in the future of the starburst
in NGC~3256.  

	It is interesting to compare our observed efficiency
of globular cluster formation in NGC~3256 to the fraction 
of mass in old stellar populations that is in globular clusters. 
The highest observed fraction of mass in globular clusters to 
total stellar mass is about $1\%$, as seen in the Galactic halo 
and the richest extragalactic globular cluster systems such as M87 
(e.g.\ Ashman \& Zepf 1998). Typical elliptical galaxies have 
ratios about five times lower, around $0.2\%$. Therefore, 
the globular cluster formation efficiency observed in NGC~3256
is more than sufficient to account for the globular cluster
systems of elliptical galaxies.
More specifically, if the mass fraction of stars that form 
in globular clusters is closer to $20\%$ over the full 
starburst/merger event, then mergers have an over-efficiency 
problem. 

	This over-efficiency problem can be alleviated in
several ways. One possibility is that the progenitor spirals 
are gas-poor, which may be true at the current
epoch but seems unlikely at high redshift when most mergers
are likely to occur. A more likely explanation is significant 
dynamical destruction of the cluster population. This
destruction is predicted by theory and required by
observation if the power-law luminosity functions of
young cluster systems are to match the lognormal luminosity
function of old globular cluster systems. 
Conversely, if the overall mass fraction
of stars that form in globular clusters is closer to $0.5\%$,
then gas-rich mergers will make about the right number
of globular clusters for moderately rich systems, and
gas-poor mergers will lead to moderately poor globular
cluster systems. The reality is likely to be somewhere
between these two extremes. It is unlikely that all globular
cluster formation in NGC~3256 will immediately cease while
the large reservoir of cold gas continues to form stars, 
as required for the overall efficiency to be $0.5\%$.
It is also unlikely that all of the remaining gas 
will form stars in a starburst in which $20\%$ of the new 
stars are formed in dense, massive clusters, as is currently 
happening in NGC~3256. Moreover, many clusters are expected
to be lost due to dynamical processes.

	The high efficiency of globular cluster formation observed
in NGC~3256 also has significant implications for theoretical
models of globular cluster formation. In particular, scenarios
in which globular clusters form as cores within proposed super 
giant molecular clouds (SGMCs) predict that only about $\sim 0.2\%$
of the mass of the cloud forms stars in dense cores, as
seen in Galactic giant molecular clouds (McLaughlin \&
Pudritz 1996, Harris \& Pudritz 1994). Thus the observed
globular cluster formation efficiency appears to pose
a problem for the SGMC scenario. A second problem is the
long timescale for the formation of SGMCs in these models
compared to the apparently rapid formation of globular clusters
in galaxy mergers and starbursts. 
The observations appear to be more consistent with models of 
globular cluster formation in which the clusters form from
highly compressed giant molecular clouds of typical mass for 
spiral galaxies. These may either originate from the progenitor 
spirals or be newly made GMCs.

\subsection{Conclusions}

	The primary conclusions of our study of HST images		
in B and I of the galaxy merger NGC~3256 are:

\noindent 1. NGC~3256 has a very large population
of compact, bright, blue objects. Many of these clusters have
estimated sizes and masses like those of Galactic globular
clusters. On this basis, we identify some fraction of these
objects as young globular clusters.

\noindent 2. The young cluster system has a broad range of
colors, but little or no correlation between color
and luminosity. This observation requires either
destruction of low mass clusters over time or a very 
young age ($\ltsim$ 20 Myr) for the young cluster system.

\noindent 3. Dynamical evolution is likely to significantly affect
the mass function of the young cluster system. If the system is 
not very young, this has already been observed. The mass-radius 
relation for the cluster population is an important constraint 
on the predictions for dynamical evolution of the cluster population.

\noindent 4. The large number of candidate young globular clusters
indicates a high efficiency of cluster formation. This
is observed across the 7 kpc $\times$ 7 kpc region studied. 
The efficiency of cluster formation in the galaxy merger
NGC~3256 is more than sufficient to account for the metal-rich 
globular cluster populations in elliptical galaxies if these form 
from gas-rich mergers.

\acknowledgments
 
	We thank Richard Larson for useful discussions, Dave Carter for 
collaborating in obtaining the AAT image of NGC~3256, and Eddie Bergeron
for making the color images frmo the HST data. We thank an anonymous 
referee for helpful suggestions.  S.E.Z. and K.M.A. acknowledge support 
for this project from NASA through grants GO-05396-94A and AR-07542-96A 
awarded by the Space Telescope Science Institute, which is operated by the 
Association of Universities for Research in Astronomy, Inc., for NASA 
under contract NAS 5-26555. SEZ also acknowledges the support of a
Hubble Fellowship and fruitful discussions with colleagues
at UC Berkeley during the early stages of this work.



\clearpage

\begin{deluxetable}{c c c c c}
\tablenum{2}
\tablecaption{Median Color of Objects as a Function B Magnitude}
\tablehead{
\colhead{$m_{B}$} & \colhead{Observed (B-I)$_{\rm{med}}$} & 
90\% confidence limits$^{a}$ & 
\colhead{Model:(B-I)$\neq$ B} & \colhead{Model:(B-I)$\propto 0.5$B}
}
\startdata
 19.25 & 0.85 & 0.81-1.10 & 0.67 & 0.36 \nl
 19.75 & 0.69 & 0.66-0.80 & 0.64 & 0.38 \nl
 20.25 & 0.67 & 0.60-0.75 & 0.64 & 0.42 \nl
 20.75 & 0.71 & 0.64-0.81 & 0.67 & 0.54 \nl
 21.25 & 0.66 & 0.57-0.69 & 0.67 & 0.65 \nl
 21.75 & 0.76 & 0.70-0.83 & 0.68 & 0.85 \nl
 22.25 & 0.82 & 0.77-0.87 & 0.73 & 1.10 \nl
 22.75 & 0.78 & 0.75-0.84 & 0.79 & 1.22 \nl
 23.25 & 0.91 & 0.87-0.94 & 0.90 & 1.36 \nl
 23.75 & 1.03 & 1.00-1.10 & 1.06 & 1.44 \nl
\enddata
\tablenotetext{a}{Confidence limits determined via bootstrap}
\end{deluxetable}


\clearpage

\begin{figure}
\caption {An R-band image of NGC~3256 obtained with the AAT.
The image is $6.6'$ on a side and is the average of a total
of 80 minutes of integration.}
\end{figure}

\begin{figure}
\caption {HST color images of NGC~3256. The first figure is
for the two WFPC2 pointings combined, while the second is
a blow-up of the central PC which is the subject of this
study. The color is created by using the image in the B-band
for the blue input, a combined B+I image for the green,
and the I-band image for the red.}
\end{figure}

\begin{figure}
\plotone{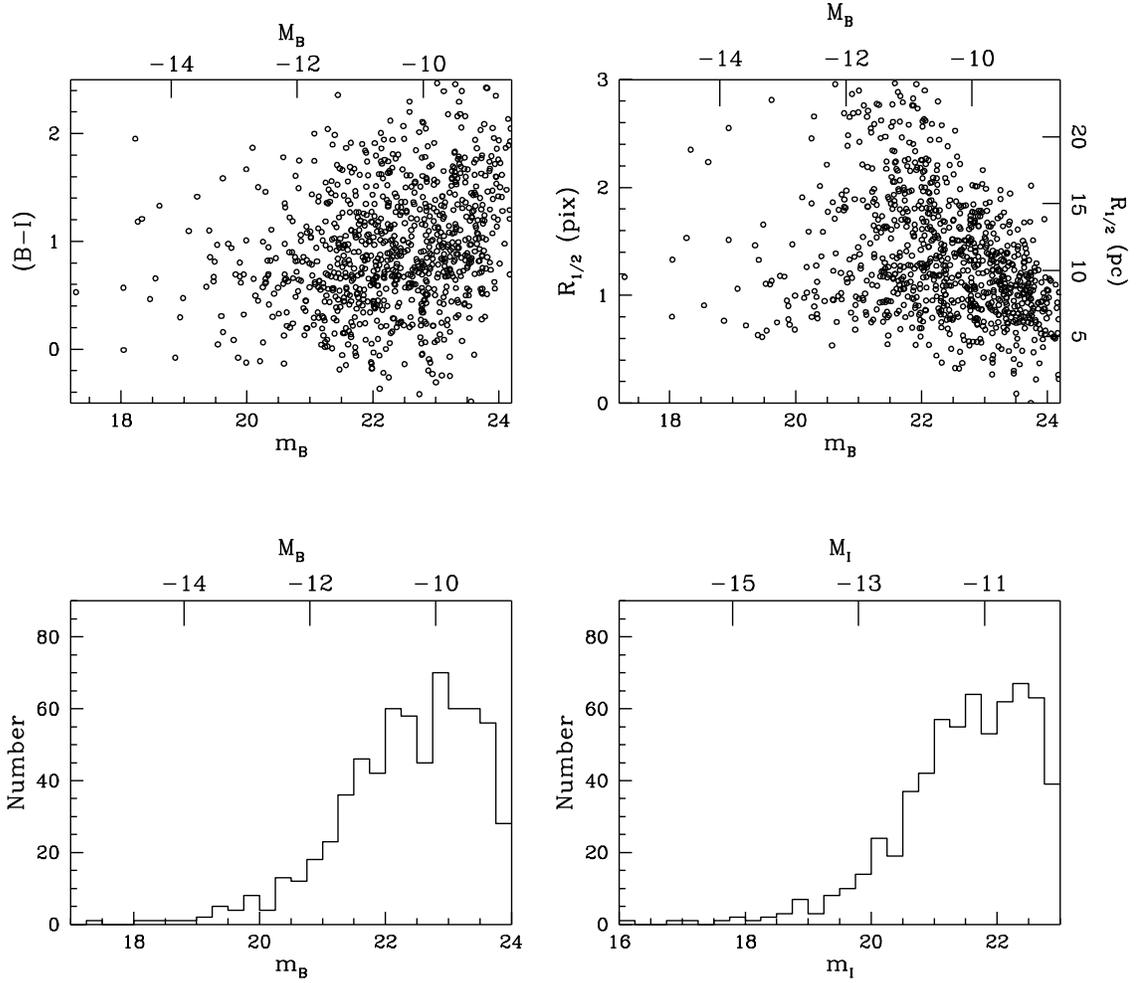}
\caption {Observed properties of the population of young globular
clusters in NGC~3256. In the upper left panel, the colors of the
clusters are plotted against their magnitudes. The upper right
panel shows the estimated cluster radii plotted against their 
magnitudes. The luminosity function of the young cluster system
is plotted in the lower panel.}
\end{figure}

\begin{figure}
\plotone{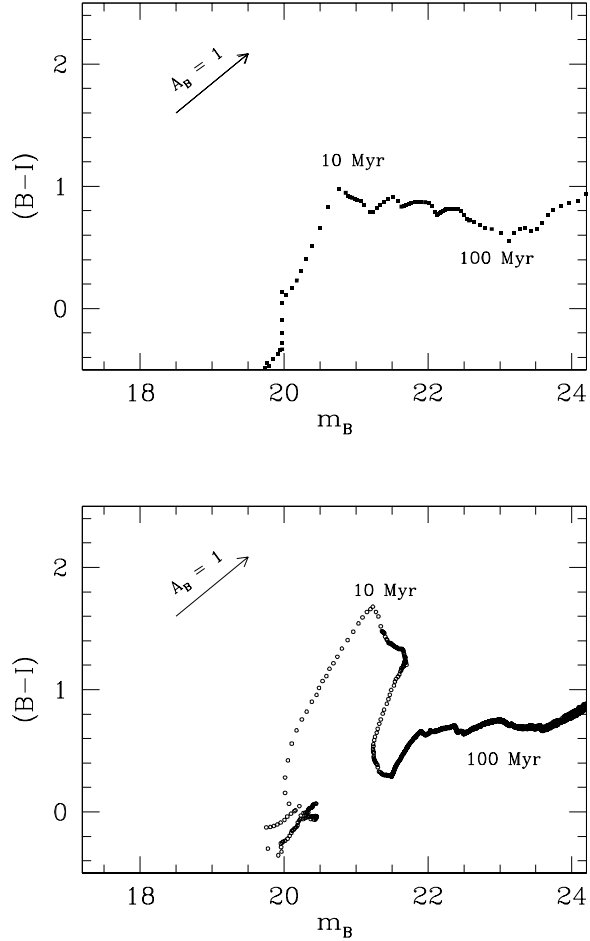}
\caption{Plots of the time evolution in color and magnitude
of a $2 \times 10^{5}$ \Msun\ star cluster. The upper panel is
based on the Bruzual \& Charlot (1998) stellar populations
models and the lower panel on the Leitherer et al.\ (1999)
models. The effect of extinction of one magnitude in B
and a Galactic extinction law 
is noted by the arrow 
in each plot, and the apparent magnitude determined 
for a distance to NGC~3256 of 37 Mpc.}
\end{figure}

\begin{figure}
\plotone{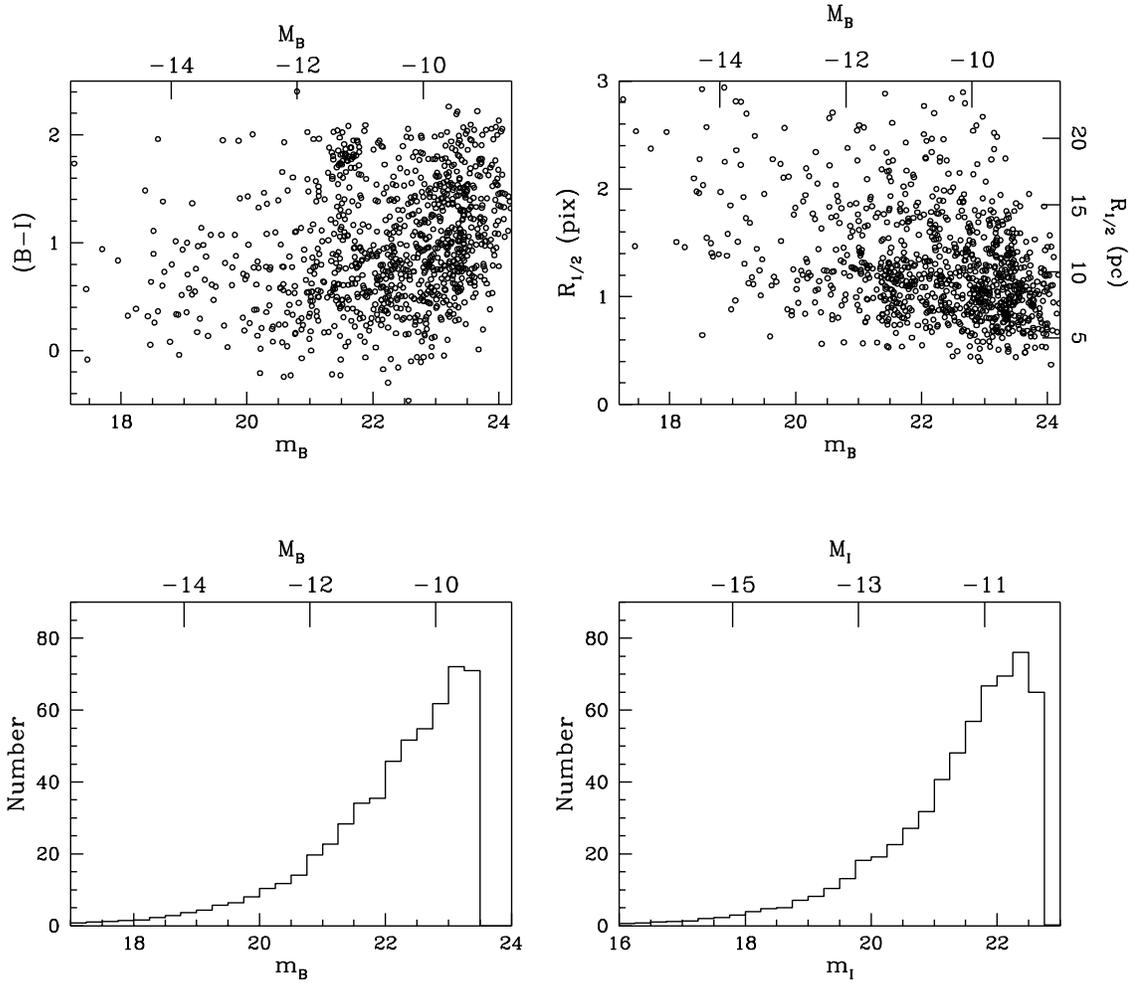}
\caption {The properties of the simulated data set which best
matches Plots of the NGC~3256 candidate young globular cluster sample
compared to a simulated dataset which is the convolution of selection 
effects with an input model with $N \propto L^{-1.8}$, 
$R_{1/2} \propto L^{0.07}$, and no correlation between $(B-I)$ and $B$.}
\end{figure}

\end{document}